\documentclass[aps,prb,preprint,groupedaddress,floatfix,showpacs]{revtex4-1}

\def\<{\langle}
\def\>{\rangle}
\def\e{\varepsilon}
\let\oldhat\hat
\renewcommand{\vec}[1]{\mathbf{#1}}
\renewcommand{\hat}[1]{\oldhat{\mathbf{#1}}}
\usepackage{graphicx}

\begin{document}

\title{Hubbard-$U$ corrected Hamiltonians for non-self-consistent 
random-phase approximation total-energy calculations: A study of ZnS, TiO$_2$, and NiO}

\author{Christopher E. Patrick}
\author{Kristian S. Thygesen}
\affiliation{Center for Atomic-Scale Materials Design (CAMD), Department of Physics,
Technical University of Denmark, DK---2800 Kongens Lyngby, Denmark}


\begin{abstract}
In non-self-consistent calculations of the total energy within the random-phase
approximation (RPA) for electronic correlation, it is necessary to choose
a single-particle Hamiltonian whose solutions
are used to construct the electronic density and non-interacting response function.
Here we investigate the effect of including a Hubbard-$U$ term in this
single-particle Hamiltonian, to better describe the on-site correlation of 3$d$
electrons in the transition metal compounds ZnS, TiO$_2$ and NiO.
We find that the RPA lattice constants
are essentially independent of $U$,
despite large changes in the underlying electronic structure.
We further demonstrate that the non-self-consistent RPA total energies of these materials 
have minima at nonzero $U$.
Our RPA calculations find the rutile phase of TiO$_2$ to be more stable than
anatase independent of $U$, a result which is consistent with experiments 
and qualitatively different to that found from calculations employing 
$U$-corrected (semi)local functionals.
However we also find that the +$U$ term cannot be used to correct
the RPA's poor description of the heat of formation of NiO.
\end{abstract}

\date{\today}

\maketitle

\section{Introduction}

Transition metal compounds (TMCs), particularly
in their nanostructured form,
find applications in a diverse range of 
technological fields including photovoltaics and photocatalysis,
magnetic storage and phosphorescent imaging.\cite{ORegan1991,
Fujishima1972,Kodama1997,Bruchez1998}
Rational optimization of TMCs at the 
nanoscale requires an atomistic,
quantum-mechanical description of these materials,
which in principle can be provided by
density-functional theory (DFT).\cite{Hohenberg1964}
Unfortunately, the most widely-used
approximations to the DFT exchange-correlation (XC) energy,
namely the local-density and generalized-gradient approximations
(LDA/GGA),
have difficulty in describing the localized $d$-electrons
of the transition metals.\cite{Ansimov1997}

This difficulty has been ascribed to the unphysical 
self-interaction experienced by the electrons within
the LDA/GGA, and a number of methods
have been proposed to overcome it.\cite{Ansimov1997}
One popular method is to
supplement the LDA/GGA XC potential with orbital-dependent $U$ terms,
designed to more accurately describe the on-site correlation of the
$d$-electrons.\cite{Ansimov1991}
Such ``Hubbard $U$'' corrections have been found to give an improved description
of the properties of TMCs like NiO.\cite{Ansimov1991,Pickett1998,Dudarev1998,Jiang2010,Cococcioni2005}

A less widely-investigated approach 
to improving the LDA/GGA description of the TMCs is to obtain the XC
energy as a combination of the ``exact'' Hartree-Fock exchange energy (EXX)
and the correlation energy calculated within the random-phase
approximation (RPA).\cite{Ren2012,Eshuis2012}
Such a scheme 
should benefit from the EXX correction
of self-interaction,\cite{Ansimov1997}
and also from the non-local and dynamical description of correlation 
provided by the RPA.
Specifically, the RPA correlation energy should capture long-range dispersive
interactions that are missing in the Hubbard $U$ corrections.\cite{Ren2012,Eshuis2012}
Recent work has demonstrated the good performance of the RPA+EXX approach 
for calculating the formation energies and relative stabilities of transition metal 
oxides.\cite{Yan2013,Jauho2015,Peng2013,Yan20132}

From the point of view of performing predictive calculations, the RPA+EXX scheme 
carries the additional advantage of being essentially parameter-free.
However, it is important to note that the most well-documented successes of this scheme---
for instance in describing non-local correlation in weakly-bonded systems,
describing chemisorption and bonding
in solids, or in the TMC examples above---were
performed non-self-consistently.\cite{Schimka2010,Olsen2011,Lu2009,Marini2006,Harl2010}
That is, the XC potential felt by the non-interacting electrons was 
not the functional derivative of the XC energy, at variance
with the standard Kohn-Sham (KS) formulation of  DFT.\cite{Kohn1965}

Although self-consistent RPA calculations have
been demonstrated, they remain a significant technical 
challenge.\cite{Nguyen2014,Klimes2014,Gruning2006,Kotani1998,Godby1988}
Therefore a key question to ask is how the choice of XC potential
in the single-particle Hamiltonian
affects the total energy calculated in a non-self-consistent RPA+EXX
scheme.
An analogy can be drawn with one-shot calculations of quasiparticle
energies within the $GW$ approximation ($G_0W_0$), where
the Green's function and screened Coulomb interaction
are usually constructed from LDA/GGA wavefunctions.\cite{Hybertsen1986}
Here it has been established that the calculated quasiparticle 
energies (e.g.\ the band gap) can depend strongly on the XC potential 
used in the single-particle Hamiltonian.\cite{Marom2012,Fuchs2007,Bruneval2013,Caruso2012}

Studies which have explored this aspect for RPA+EXX total energy calculations 
have usually focused
on the differences between LDA and GGA or on the
effect of including Hartree-Fock exchange.\cite{Jiang2007, Gruneis2009,
Harl2010,Toulouse2010,Angyan2011,Olsen20132}
In most cases, the initial choice of XC potential has been found to play only
a minor role; a notable exception is the study of cerium in Ref~\citenum{Casadei2012},
and of molecular dissociation in Refs.~\citenum{Caruso2013}~and~\citenum{Hellgren2015}.
However, for TMCs it is natural to investigate the 
effect on the RPA+EXX total energy of adding 
Hubbard $U$ corrections to the XC potential.
Since such corrections can significantly change the 
character of the single-particle wavefunctions and
their energy eigenvalues, one might expect to observe 
a dependence of the RPA+EXX total energy on the
parameter $U$.
Indeed, one might even hope that including Hubbard $U$ corrections
in the XC potential might improve the quality of the subsequent
RPA+EXX calculation, if the resulting single-particle wavefunctions
are closer to the exact KS form.\cite{Peng2013}
On the other hand, it is important to note that the orbital-dependent  
Hubbard $U$ corrections are non-local, and that
the RPA correlation energy is strictly non-variational with respect
to all possible non-local XC potentials.\cite{IsmailBeigi2010}

Motivated by these considerations, we have performed
a systematic study of the effects of Hubbard $U$ corrections
on the non-self-consistent RPA+EXX total energy of TMCs.
We present results for ZnS, TiO$_2$
and NiO which, in terms of their 3$d$ states, display  progressively more
complex electronic structure.
From the total energies we obtain lattice constants as a function
of the $U$ parameter within the RPA approximation for the correlation energy,
and compare the results to non-self-consistent EXX,
self-consistent GGA+$U$ or LDA+$U$, and experiment.
We also consider the energetics of the technologically-important
TiO$_2$ polymorphs of anatase and rutile, and the heats of formation
of TiO$_2$ and NiO.

The rest of our paper is organized as follows.
In Section~\ref{sec.methods} we outline the theory
of the non-self-consistent RPA+EXX scheme and describe our computational approach.
In Sections~\ref{sec.ZnS}, \ref{sec.TiO2} and \ref{sec.NiO} we
present our results for ZnS, TiO$_2$ and NiO, including our
calculations of the phase stability of TiO$_2$ in Section~\ref{sec.TiO2_relstab}.
We provide a detailed analysis of the $U$-dependence
of the total energy in Sections~\ref{sec.U_energies}~and~\ref{sec.U_min}, and
consider the oxide heats of formation in Section~\ref{sec.heat_form}.
We present our conclusions in Section~\ref{sec.Conclusions}.

\section{Theory and computational methodology}
\label{sec.methods}

\subsection{Non-self-consistent RPA total energy}

We consider the ground-state total energy $E_\mathrm{Tot}$
of a system of electrons and
nuclei, treating the nuclei as classical, stationary
particles.
Within the adiabatic-connection fluctuation-dissipation
formulation of DFT,\cite{Langreth1977,Gunnarsson1976,Ren2012,Eshuis2012} 
$E_\mathrm{Tot}$ is decomposed as
\begin{equation}
E_\mathrm{Tot} = E_0 + E_\mathrm{X} + E_\mathrm{C}.
\label{eq.etot}
\end{equation}
The quantity $E_0$ appearing in equation~\ref{eq.etot}
is the total energy neglecting exchange and correlation,
given by
\begin{equation}
E_0 = T_s\left[\{\psi\} \right] + E_{Ie}[\rho] + E_\mathrm{Har}[\rho] + E_{II},
\label{eq.E0}
\end{equation}
where $\{\psi\}$ denotes
the set of single-particle wavefunctions obtained from solving
\begin{equation}
H^0 |\psi_{\nu\sigma}\> = \e_{\nu\sigma} |\psi_{\nu\sigma}\>,
\label{eq.HKS}
\end{equation}
where $H^0$ is a single-particle Hamiltonian (Section~\ref{sec.H0}).
The electronic density $\rho$ is constructed as $\sum_{\nu\sigma} f_{\nu\sigma}|\psi_{\nu\sigma}|^2$,
where $f_{\nu\sigma}$ gives the occupation number of the state.
For crystalline systems $\nu$ 
is a composite index labelling band index and wavevector, and $\sigma$
is a spin index (here we assume collinear spin polarization).
$T_s$ gives the kinetic
energy of the single-particle wavefunctions, and $E_{Ie}$, $E_\mathrm{Har}$
and $E_{II}$ give the electron-nuclear, electron-electron,
and nuclear-nuclear electrostatic  interaction energies.

The exchange energy $E_\mathrm{X}$ is obtained as
\begin{eqnarray}
E_\mathrm{X} &=& 
-\frac{1}{2}
\sum_{\nu_1,\nu_2,\sigma} f_{\nu_1\sigma}f_{\nu_2\sigma} \times \nonumber \\
&&\int d\vec{r} \int d\vec{r'}
\frac{ \psi_{\nu_1\sigma}(\vec{r}) \psi_{\nu_2\sigma}^*(\vec{r}) \psi_{\nu_2\sigma}(\vec{r'}) \psi_{\nu_1\sigma}^*(\vec{r'})}
{ |\vec{r}-\vec{r'}|}, 
\label{eq.Ex}
\end{eqnarray}
(Hartree units are used throughout),
and the correlation energy $E_\mathrm{C}$ is expressed as
\begin{eqnarray}
E_\mathrm{C} = -\frac{1}{2\pi} \int_0^1&& d\lambda \int_0^\infty ds
\int d\vec{r} \int d\vec{r'} \times \nonumber \\
&&\frac{ \chi^\lambda(\vec{r},\vec{r'};is) - \chi_\mathrm{KS}(\vec{r},\vec{r'};is)}
{ |\vec{r}-\vec{r'}|}
\label{eq.Ec_full}
\end{eqnarray}
where $s$ is a real number representing an imaginary frequency, $\omega = is$.
$\lambda$ is a coupling constant taking values between 0 and 1 which controls the 
strength of the Coulomb interaction along the adiabatic connection, and defines
a Hamiltonian $H^\lambda$ whose solution
yields the exact ground-state electronic density for all $\lambda$.

The response functions $\chi$ appearing in the integrand of equation~\ref{eq.Ec_full} 
are related through an integral equation.\cite{Petersilka1996}
Within the RPA this equation can be inverted to give
$\chi^\lambda_\mathrm{RPA}(\omega) = [1 - \lambda \chi_\mathrm{KS}(\omega) v_\mathrm{C} ]^{-1}  \chi_\mathrm{KS}(\omega)$,
where $v_\mathrm{C}$ is the Coulomb interaction.
Integrating over the coupling constant in equation~\ref{eq.Ec_full} and
expanding the response function in a plane-wave basis yields
the RPA correlation energy,\cite{Ren2012,Eshuis2012}
\begin{eqnarray}
E_\mathrm{C}^\mathrm{RPA} = \frac{1}{2\pi} \sum_\vec{q} \int_0^\infty ds \ 
\mathrm{Tr}&&\left[\ln\{ 1 - v_\mathrm{C}(\vec{q})\chi_\mathrm{KS}(\vec{q},is)\} \right. \nonumber \\
&&\left. + v_\mathrm{C}(\vec{q})\chi_\mathrm{KS}(\vec{q},is)\right]
\label{eq.Ec}
\end{eqnarray}
where $\vec{q}$ is a wavevector in the first Brillouin zone,
and the response function is a matrix in the reciprocal lattice
vectors $\vec{G}$ and $\vec{G'}$, with elements given by\cite{Yan2011}
\begin{eqnarray}
\chi_\mathrm{KS}^{\vec{G}\vec{G'}}(\vec{q},is) &=& \frac{1}{\Omega}
\sum_{\vec{k}nn'\sigma}(f_{n\vec{k}\sigma} - f_{n'\vec{k}+\vec{q} \sigma}) \times \nonumber \\
&&
\frac{
n^\sigma_{n\vec{k},n'\vec{k}+\vec{q}}(\vec{G})
n^{\sigma *}_{n\vec{k},n'\vec{k}+\vec{q}}(\vec{G'})
}
{is + \e_{n\vec{k}\sigma} - \e_{n'\vec{k}+\vec{q}\sigma}}.
\label{eq.chiks}
\end{eqnarray}
$\Omega$ is the volume of the primitive unit cell, and 
the pair density $n^\sigma_{n\vec{k},n'\vec{k}+\vec{q}}(\vec{G})
=
\<\psi_{n\vec{k}\sigma}| e^{-i(\vec{q}+\vec{G})\cdot\vec{r}} | \psi_{n'\vec{k}+\vec{q}\sigma}\>$.

Setting $E_\mathrm{C}$ to $E_\mathrm{C}^\mathrm{RPA}$ in equation~\ref{eq.etot} completes
our prescription for a calculation of the RPA total energy $E^\mathrm{RPA}_\mathrm{Tot}$.
The density and response function are constructed
from the set of wavefunctions which solve equation~\ref{eq.HKS},
and the separate contributions to $E^\mathrm{RPA}_\mathrm{Tot}$ are evaluated from
equations~\ref{eq.E0}, \ref{eq.Ex} and \ref{eq.Ec}, i.e.\
\begin{equation}
E^\mathrm{RPA}_\mathrm{Tot} = E_0 + E_\mathrm{X} + E_\mathrm{C}^\mathrm{RPA}.
\label{eq.etotRPA}
\end{equation}
For comparison we also consider the non-self-consistent total energy only including
the exact exchange (EXX) contribution,
\begin{equation}
E^\mathrm{EXX}_\mathrm{Tot} =  E_0 + E_\mathrm{X}.
\label{eq.etotEXX}
\end{equation}
In passing we point out that by defining $E_0$ as in equation~\ref{eq.E0}
we remove the need to include double-counting corrections in 
equations~\ref{eq.etotRPA}~and~\ref{eq.etotEXX}
(to be contrasted with e.g.\  equations 7 and 27 of Ref.~\citenum{Ren2012}).

\subsection{Single-particle Hamiltonian}
\label{sec.H0}

The procedure outlined in the previous section of
calculating $E^\mathrm{RPA}_\mathrm{Tot}$
leads to an ambiguity in the definition of the
single-particle Hamiltonian $H^0$.
As mentioned above, the adiabatic connection depends on
the exact density being recovered for all values
of $\lambda$.
Equation~\ref{eq.HKS} corresponds to $\lambda = 0$,
thus identifying $H^\mathrm{0}$ as the single-particle Hamiltonian
which yields the exact density of the system of interacting
electrons, i.e.\ the Kohn-Sham (KS) Hamiltonian with
the exact exchange-correlation (XC) potential $V_\mathrm{XC}$.\cite{Kohn1965}
One approach therefore would be to 
use the solutions of this Hamiltonian (equation~\ref{eq.HKS}) to compute
the contributions $E_0 + E_\mathrm{X}$ in equation~\ref{eq.etot}, 
independent of any subsequent approximation
used to compute $E_\mathrm{C}$ (e.g.\ the RPA).
However such an approach relies on having the exact $V_\mathrm{XC}$, 
which is unfortunately not known.

An alternative approach is to treat the combined quantity 
$E_\mathrm{XC}^\mathrm{RPA} = E_\mathrm{X} + E_\mathrm{C}^\mathrm{RPA}$
as an orbital-dependent XC-functional, and use
a Kohn-Sham Hamiltonian in equation~\ref{eq.HKS} with
an XC potential constructed as a functional derivative,
$V_\mathrm{XC}^\mathrm{scRPA} = \delta E_\mathrm{XC}^\mathrm{RPA}/\delta \rho$.
This self-consistent (sc) RPA scheme ensures compatibility between the total
energy functional and XC potential.\cite{Nguyen2014,Klimes2014,Gruning2006,Kotani1998,Godby1988}
In contrast to the non-self-consistent case, in this scheme the RPA is being used
to determine $H^0$ and thus $E_0 + E_\mathrm{X}$.
Therefore the scRPA
scheme can no longer be considered as an approximation
to $E_\mathrm{C}$ alone.
Of course since
$E_\mathrm{XC}^\mathrm{RPA}$ is nonlocal and energy-dependent,
it may also be hoped that 
$V_\mathrm{XC}^\mathrm{scRPA}$ might represent a better approximation
to the unknown, exact $V_\mathrm{XC}$
than simpler functionals like the LDA/GGA.

In this work we focus on the first (non-self-consistent) approach,
and approximate the exact $V_\mathrm{XC}$
with one chosen from the class of functionals which include a 
Hubbard $U$ term.
Specifically we supplement standard LDA/GGA XC-functionals
with the correction
derived in Ref.~\citenum{Dudarev1998},
\begin{equation}
\Delta E_U = \frac{U}{2} \sum_a \mathrm{Tr} (\rho^a - \rho^a\rho^a).
\label{eq.Ucorr}
\end{equation}
The density matrices $\rho^a$ describe the occupation of localized
orbitals on atom $a$, and $U$ controls the strength of the on-site
Coulomb interaction incorporating both Hartree ($U_H$) and exchange ($J$)
contributions, $U$ = $U_H$ $-$ $J$.\cite{Dudarev1998}
The LDA/GGA+$U$ XC potential $V^U_\mathrm{XC}$ is constructed using equation~\ref{eq.Ucorr}
following the scheme described in Refs.~\citenum{Enkovaara2010}~and~\citenum{Rohrbach2004},
with the $d$-projectors located on Zn, Ti and Ni atoms 
defining the density matrices appearing in equation~\ref{eq.Ucorr}.\cite{Enkovaara2010}
The single particle Hamiltonian used in equation~\ref{eq.HKS} is thus
\begin{equation}
H^0(U) = [T + V_{Ie} + V_\mathrm{Har}] + V^U_\mathrm{XC},
\label{eq.spH}
\end{equation}
where the operators in the square brackets are obtained as the functional
derivative of $E_0$ (equation~\ref{eq.E0}).
We emphasize that $U$ is considered a free parameter which, for
a given choice of LDA or GGA, completely
determines $H^0$ (and thus $E^\mathrm{RPA}_\mathrm{Tot}$) through
equation~\ref{eq.spH}.

\subsection{Computational details}

All calculations were performed within the projected-augmented
wave (PAW) formalism\cite{Blochl1994} of DFT\cite{Hohenberg1964,Kohn1965}
as implemented in the \texttt{GPAW} code.\cite{Enkovaara2010}
The core-valence interaction was described using the 
0.9.11271 \texttt{GPAW} datasets, which always treat
the 4$s$ and 3$d$ shells of the transition metals as valence states,
and further explicitly include the 3$s$ and 3$p$ shells
for Ti and 3$p$ shell for Ni.
Exchange and correlation effects were described either
within the LDA\cite{Perdew1992} or GGA (the PBE XC-functional)\cite{Perdew1996}
with the Hubbard $U$ correction scheme described 
above.\cite{Dudarev1998, Rohrbach2004, Enkovaara2010}

\begin{table}
\caption{
\label{tab.kpts} 
Size of $\Gamma$-centred Monkhorst-Pack grids\cite{Monkhorst1976} used in the calculation of
EXX total energy and RPA correlation energy for each material.
}
\begin{tabular}{lll}
\hline
\hline
                    &  $E_0 + E_\mathrm{X}$    & $E_\mathrm{C}^\mathrm{RPA}$\\
\hline
ZnS                 &  10$\times$10$\times$10  & 6$\times$6$\times$6 \\
TiO$_2$ (rutile)    &  6$\times$6$\times$8     & 4$\times$4$\times$6 \\
TiO$_2$ (anatase)   &  8$\times$8$\times$4     & 6$\times$6$\times$4 \\
NiO\footnotemark[1] &  8$\times$8$\times$4     & 8$\times$8$\times$4 \\
Ti                  &  22$\times$22$\times$22  & 12$\times$12$\times$12 \\
Ni                  &  22$\times$22$\times$22  & 14$\times$14$\times$14 \\
\hline
\hline
\footnotetext[1]{1$\times$1$\times$2 supercell used to describe antiferromagnetic unit cell}
\end{tabular}
\end{table}
The electronic wavefunctions were expanded in plane
waves up to a maximum energy of 80~Ry.
The wavefunctions were sampled on the $\Gamma$-centred Monkhorst-Pack\cite{Monkhorst1976}
grids listed in Table~\ref{tab.kpts}.
For the metals, the electronic occupations were modelled with a Fermi-Dirac
distribution of width 0.01~eV.
The small-wavevector divergence of the Coulomb interaction
was handled with the Wigner-Seitz truncation scheme of
Ref.~\citenum{Sundararaman2013} when calculating the exchange
energy, and with the perturbation theory approach described
in Ref.~\citenum{Yan2011} when calculating the correlation energy.

The response function $\chi_\mathrm{KS}$ was expanded in plane
waves up to a maximum energy $E_\mathrm{cut}$ of 30~Ry.
Following previous studies\cite{Harl2010,Olsen20132} we set the number of unoccupied
bands used in equation~\ref{eq.chiks} equal to the number of plane waves
used to describe $\chi_\mathrm{KS}$, and extrapolated the
results obtained at finite $E_\mathrm{cut}$  (20--30~Ry) 
to the basis set limit using the power law expression
$E_\mathrm{C}^\mathrm{RPA}(E_\mathrm{cut}) = E_\mathrm{C}^\mathrm{RPA}(\infty) + AE_\mathrm{cut}^{-3/2}$.
The frequency integration in equation~\ref{eq.Ec}
was performed numerically within the scheme described in Ref.~\citenum{Olsen20132}.

The geometry optimizations of ZnS and NiO were performed by calculating the
total energy for seven lattice parameters, spanning $\pm$7\% around the experimental
value, and fitting the calculated energies to the Birch-Murnaghan equation of state.\cite{Birch1947}
To optimize the geometry of TiO$_2$ (which is a function of three independent parameters),
we fixed two of the parameters at their previous ``best'' values and calculated the
energy as a function of the third, which we varied by $\pm$7\% around the experimental value.
After fitting a polynomial to the total energy we obtained a new ``best'' value 
for this parameter.
We repeatedly cycled through all the parameters until no change was observed between iterations.
For consistency we used this procedure for the EXX, RPA+EXX and PBE+$U$ calculations,
even though geometry optimization for the latter can be achieved more easily
using the stress theorem.\cite{Nielsen1985}

For the heat of formation calculations, we modeled the O$_2$ molecule 
in its triplet state
with a fixed bond length of 1.21~\AA .
We used periodic simulation cells and sampled the wavefunctions at the $\Gamma$-point.
For the calculation of $E_0 + E_\mathrm{X}$ we used a simulation cell of size 12$\times$12$\times$13~\AA$^3$,
and a cell of size of 6$\times$6$\times$7~\AA$^3$ \ 
for $E_\mathrm{C}^\mathrm{RPA}$.
\section{Results and discussion}
\subsection{ZnS}
\label{sec.ZnS}
\subsubsection{Electronic structure}
\begin{figure}
\includegraphics{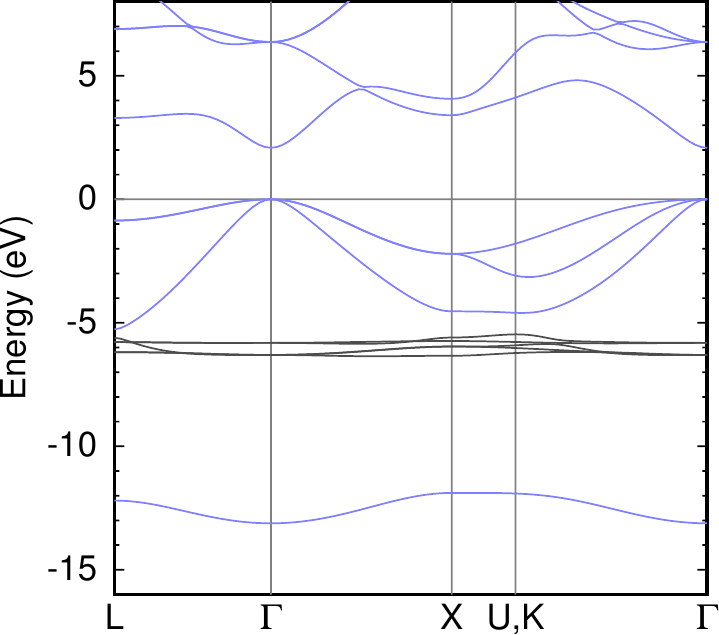}
\caption{(color online)
Electronic bandstructure of sphalerite ZnS calculated
at the experimentally-measured lattice constant\cite{Hotje2003} 
using the PBE XC-functional
(no $U$ correction).
The energy zero has been set to the top of the valence band.
The 3$d$ band originating from the Zn atoms is located at -6~eV and
highlighted in black.
\label{fig.ZnS_bs}
}
\end{figure}
We begin our study by considering ZnS
in its sphalerite form (zinc blende, F$\overline{4}3m$).
The electronic bandstructure calculated using the 
PBE XC-functional is shown in Fig.~\ref{fig.ZnS_bs}.
As found in numerous previous LDA/GGA calculations\cite{Miyake2006,Rinke2005,
Vogel1996,Karazhanov2007,Kotani2002,Shishkin2007}
the filled Zn-3$d$ shells form a narrow band at 6~eV below
the valence band edge. 
This 3$d$ band is also observed in valence
photoemission experiments, but at a 
larger binding energy of 9~eV.\cite{Ley1974,Weidemann1992}

Adding a Hubbard $U$ correction to the PBE XC-functional 
shifts the 3$d$ band to larger binding energy, with the
magnitude of the shift depending linearly on $U$.
We find the 3$d$ band position to coincide with the experimental binding
energy when $U\approx 8$~eV.
This value is consistent with two previous LDA+$U$ studies\cite{Jiang2010,Miyake2006}
which required $(U_H-J)$ values of 9 and 7~eV to shift the 3$d$ band 
to the experimentally-observed position.
Like these studies\cite{Jiang2010,Miyake2006} we also 
observe that the band gap depends weakly on
$U$, increasing from 2.1 to 2.6~eV when $U$ is varied from 0 to 10~eV.

\subsubsection{Atomistic structure}
\begin{figure}
\includegraphics{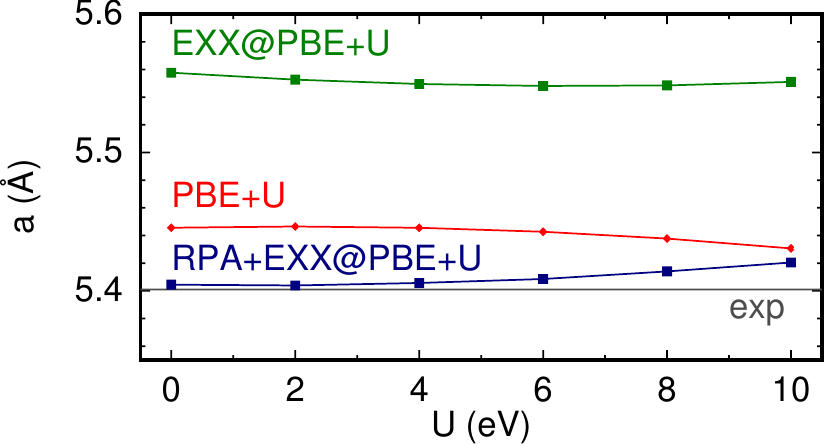}
\caption{(color online)
Lattice constant $a$ of sphalerite ZnS calculated using total energies
obtained self-consistently with the PBE XC-functional and Hubbard $U$ correction
(red), or non-self-consistently starting from PBE+$U$ wavefunctions
and eigenvalues including exact exchange without (green) and with (blue)
the RPA correlation energy, as a function of $U$.
The lines are guides to the eye.
The lattice constant measured in Ref.~\citenum{Hotje2003}
is shown as a gray horizontal line.
\label{fig.ZnS_structure}
}
\end{figure}
In Fig.~\ref{fig.ZnS_structure} we show the equilibrium lattice constant
calculated as a function of $U$, either at the PBE+$U$ level
or from the non-self-consistent RPA+EXX and EXX total energies calculated from
equations~\ref{eq.etotRPA}~and~\ref{eq.etotEXX}.
We compare our calculations to the value of 5.401~\AA \ measured
from X-ray diffraction\cite{Hotje2003} (horizontal line in Fig.~\ref{fig.ZnS_structure}).
Considering the PBE+$U$ calculations first (red line), at $U$=0~eV we observe a lattice constant
which is 0.8\% larger than the reported experimental value.
This difference is maintained over the $U$ range of 0--6~eV and then slightly
decreases, to 0.5\% for $U$=10~eV.
This magnitude of variation is rather small compared to the other materials discussed below,
which we attribute to the energetic separation of the $3d$ bands.
The other bands, lying 0--5 and 12--13~eV below the valence band maximum (VBM), have
predominantly S-3$p$/Zn-4$s$ and S-3$s$ character respectively.
We note that for large values of $U$ the Zn-3$d$ band is pushed down in energy sufficiently to begin
to hybridize with the S-$3s$ states.
Indeed fixing the lattice constant and monitoring the band character as a function of $U$ 
shows a rapid increase in the Zn-3$d$ contribution to the S-$3s$ band for  values
of $U \geq$ 8~eV.

Next considering the lattice constants obtained from the non-self-consistent exact
exchange energy $E^\mathrm{EXX}_\mathrm{Tot}$ (green line in Fig.~\ref{fig.ZnS_structure}),
we find a value 2.9\% larger than experiment at $U$=0~eV.
This difference is varies by less that 0.2\% over the full range of $U$ values.
Although EXX lattice constants are often overestimated with respect to experiment,\cite{Harl2010}
2.9\% is somewhat larger than the mean absolute error of 1.2\% obtained in Ref.~\citenum{Harl2010}
for a test set of 20 semiconductors, which included several zinc blende structures.
The correlation contribution to the total energy should therefore be considered 
particularly important to the bonding of ZnS.

Finally we consider lattice constants obtained after 
adding the non-self-consistent RPA correlation energy to the EXX energy,
$E^\mathrm{RPA}_\mathrm{Tot}$ (blue line).
Here we find lattice constants very close to the experimental value: 5.40 and 
5.42~\AA \ at $U$=0~and~10~eV,
corresponding to increases of $<$0.1\% and 0.4\% respectively.
The variation of lattice constant with $U$ displays the opposite trend to the PBE+$U$
calculations; in fact, the behavior is almost a perfect mirror image.
That is, the XC-interaction which favors increased bonding at high $U$ within
the PBE+$U$ approximation is not present within the RPA description of the correlation
energy. 

Overall, our results show that the calculated lattice constant of sphalerite ZnS is
somewhat insensitive to the value of $U$ used in $H^0$, at all levels of theory.
The fact that the Zn-3$d$ states are already fully occupied and located deep
below the VBM for $U$=0~eV means that adding a $U$ correction to
these orbitals has a minimal effect on the ground-state electron density.

\subsection{TiO$_2$}
\label{sec.TiO2}

\subsubsection{Electronic structure}
\begin{figure}
\includegraphics{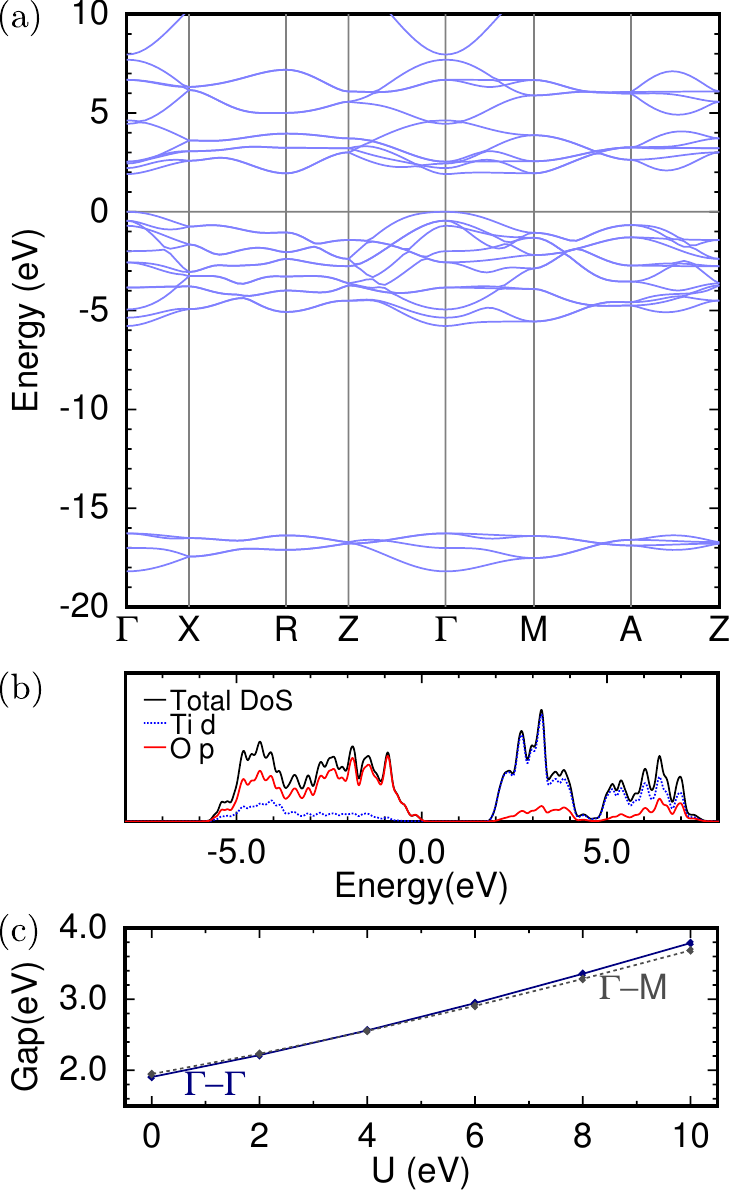}
\caption{(color online)
(a) Electronic bandstructure of rutile TiO$_2$ calculated
at the experimental structure\cite{Burdett1987} using the PBE XC-functional
(c.f.\ Fig.~\ref{fig.ZnS_bs}).
(b) Density-of-states (DoS) of rutile around the valence and conduction band
(black line) projected onto the Ti-$d$ and O-$p$ PAW projector functions
(blue and red lines).
(c) Evolution of the energy gaps with Hubbard $U$ correction applied
to Ti-3$d$ states, corresponding to the direct transition
at the $\Gamma$ point ($\Gamma-\Gamma$, blue solid line) and the indirect
transition ($\Gamma-M$, gray dotted line).
\label{fig.TiO2_bs_gap}
}
\end{figure}

We now consider TiO$_2$, a material where the 3$d$ shell is largely unoccupied.
The most naturally-abundant forms of TiO$_2$ are the rutile (P$4_2$/$mnm$)
and anatase (I$4_1$/$amd$) polymorphs.\cite{Muscat2002}
We begin by focusing on rutile TiO$_2$, and calculate the electronic
bandstructure and projected density-of-states (PDoS)
at the PBE level using experimental structural parameters.\cite{Burdett1987}
The results are shown in Figs.~\ref{fig.TiO2_bs_gap}(a)~and(b).
The valence and conduction bands are formed from a mix of O-2$p$ and
Ti-3$d$ states, with O-2$p$ dominating the valence band and vice versa.
The Ti-3$d$ states in the conduction band are further split by the crystal field  into $t_{2g}$ and
$e_g$ subbands, over the energy region 2--4.5~eV and 4.5--7.5~eV above the VBM.
The O-2$s$ states lie far (17~eV) below the VBM.
These electronic structure features have been observed and discussed
in numerous other works.\cite{Goodenough1971,Glassford1992,
Labat2007,Chiodo2010,Kang2010,Dompablo2011}

The effect of including a Hubbard $U$ correction to the Ti-3$d$ states
is to reduce the hybridization with the O-2$p$ orbitals
in the conduction and valence bands, and to push the $t_{2g}$ subband 
up in energy.\cite{Dompablo2011}
The latter phenomenon leads to a strong dependence of the fundamental
gap on $U$,\cite{Dompablo2011,Patrick2012} 
illustrated in Fig.~\ref{fig.TiO2_bs_gap}(c).
The direct gap at the $\Gamma$ point increases by almost 2~eV over
the $U$-range of 0--10~eV, 4 times larger than observed for ZnS.
As also shown in Fig.~\ref{fig.TiO2_bs_gap}(c) the small
difference between the direct gap at $\Gamma$ and the
$\Gamma$--$M$ transition (0.04~eV at $U$=0~eV) reduces to zero
at $U$=4~eV, such that the nature of the fundamental gap
changes from direct to indirect for $U\geq$4~eV.\cite{Patrick2012}

\subsubsection{Atomistic structure}
\begin{figure}
\includegraphics{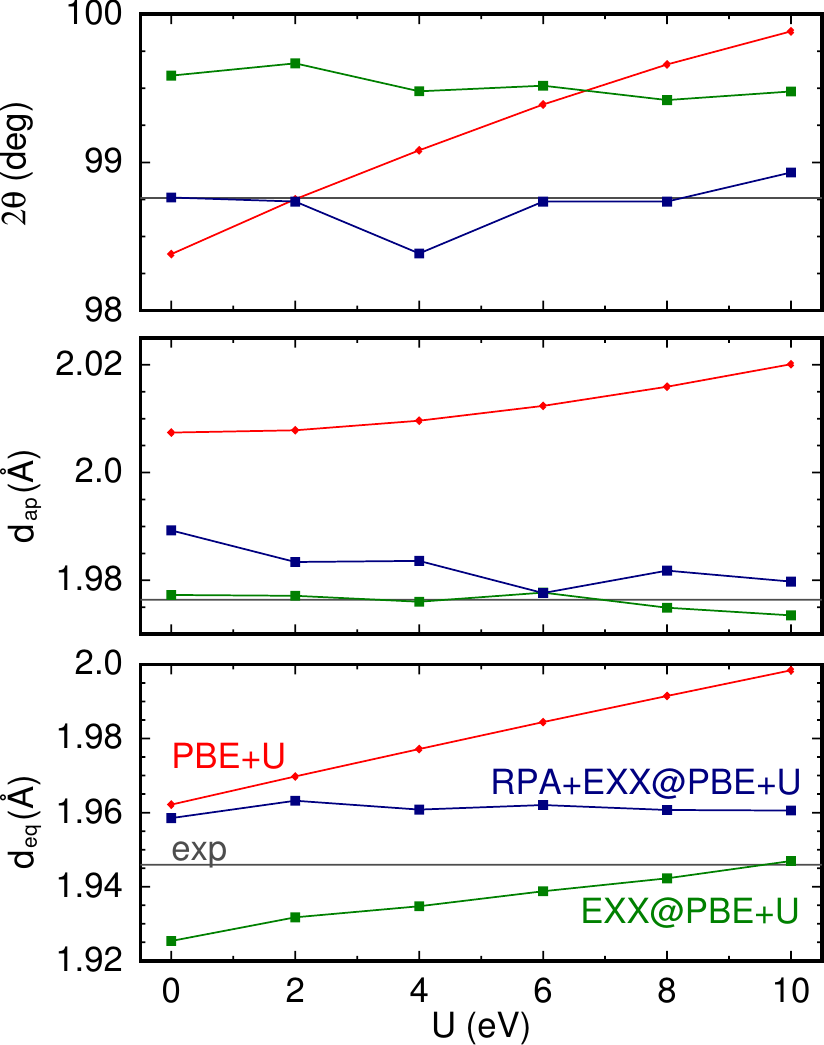}
\caption{(color online)
Structural parameters of rutile TiO$_2$ calculated under different
approximations (c.f.\ Fig.~\ref{fig.ZnS_structure} for labels).
The meaning of the three parameters is given in the main text.
The experimental structural parameters (gray horizontal lines) 
were measured in Ref.~\citenum{Burdett1987}.
\label{fig.TiO2_structure}
}
\end{figure}

The structure of rutile TiO$_2$ is fully specified by the lattice
parameters $a$ and $c$ and a dimensionless internal parameter $u$.
Equivalently the structure may be described\cite{Fahmi1993}
in terms of distorted TiO$_6$ octahedra characterized by apical
and equatorial bond lengths ($d_\mathrm{ap}$ and $d_\mathrm{eq}$)
and an angle $\theta$, where $2\theta$ is the smallest Ti--O--Ti
angle in a given OTi$_3$ planar unit.
The two parameter sets are related through:
\begin{eqnarray}
d_\mathrm{ap} &=& u a \sqrt{2} \label{eq.dap}\\
d_\mathrm{eq} &=& \frac{a}{2} \sqrt{\left(\frac{c}{a}\right)^2 + 8\left(\frac{1}{2} -u\right)^2} \\
\cos 2\theta  &=& \frac{2a^2\left(u - \frac{1}{2}\right)^2 - \frac{c^2}{4}}{2a^2\left(u - \frac{1}{2}\right)^2 + \frac{c^2}{4}} \label{eq.theta}
\end{eqnarray}
The inversions of equations~\ref{eq.dap}--\ref{eq.theta}
are given in Ref.~\citenum{Fahmi1993}.

Figure~\ref{fig.TiO2_structure} shows the calculated values 
of the parameters $d_\mathrm{ap}$, $d_\mathrm{eq}$ and $2\theta$ 
as a function of $U$ using PBE+$U$,
and non-self-consistent EXX and RPA+EXX total energies.
We also show the structural parameters obtained in the
neutron diffraction experiments of Ref.~\citenum{Burdett1987},
corresponding to $a=4.587$~\AA, $c=2.954$~\AA \ and $u=0.3047$.
Considering the PBE+$U$ data first, there is a strong dependence
of the three parameters on the value of $U$ used.
$d_\mathrm{eq}$ and $d_\mathrm{ap}$ increase by 1.8\%~and~1.3\% between $U$=0~and~10~eV,
which is a much larger change than the 0.3\% decrease in Zn-S bond length observed
for ZnS.
A simple explanation for the observed lengthening of bonds is that
the $U$ correction makes the orbitals more atomic-like, reducing
the hybridization shown in Fig.~\ref{fig.TiO2_bs_gap}(b) and thus weakening the bonding.\cite{Dompablo2011}
The increased $U$ also drives $2\theta$ away from 90$^\circ$ and towards
120$^\circ$, which as noted in Ref.~\citenum{Fahmi1993} is its optimal value 
from the point of view of the planar threefold co-ordination of the O atoms;
that is, the importance of the O atoms to the bonding increases
with $U$.

Moving onto the EXX calculations, we see that $d_\mathrm{ap}$ and
$2\theta$ are effectively independent of $U$.
$d_\mathrm{ap}$ is particularly close ($<0.1$\%)
to the experimental value, whilst $2\theta$ is overestimated
by 0.8\%.
However, $d_\mathrm{eq}$ displays a monotonic $U$-dependence,
with deviation from the experimental value varying
from -1\% to $<0.1$\% for $U$ between 0~and~10~eV.
We note that the variation of $E^\mathrm{EXX}_\mathrm{Tot}$ with
$U$ can only be due to the change in the shape of the occupied
orbitals, which determines $E_0$ and $E_\mathrm{X}$.
We also note that the EXX structural parameters are
closer to experiment than found for ZnS.
This result is consistent with Refs.~\citenum{Muscat2002}~and~\citenum{Labat2007},
which found the structures calculated within the Hartree-Fock approximation
(self-consistent EXX) to be close to experimental values.

Given the apparent sensitivity of the EXX calculations of $d_\mathrm{eq}$
to the $U$ value used,
we might also expect the RPA+EXX structural parameters to exhibit
a $U$-dependence.
In particular, since the denominator of $\chi_\mathrm{KS}$
in equation~\ref{eq.chiks} consists of energy differences between
occupied and unoccupied states, the increase in band gap
shown in Fig.~\ref{fig.TiO2_bs_gap}(c) should introduce an
additional coupling between $H^0(U)$ and $E^\mathrm{RPA}_\mathrm{Tot}$.
What we observe however is that the RPA+EXX calculations are 
rather insensitive to the value of $U$ used (blue lines in Fig.~\ref{fig.TiO2_structure}).
Furthermore, the calculated structures are close to experiment;
at $U$=0~eV we find values of 4.616~\AA, 2.973~\AA \ and 0.3047 for $a$, $c$ and $u$,
which are all within 0.7\% of experiment.
There is noticable noise in the data, particularly for the calculated $2\theta$,
which reflects the difficulty in fitting the RPA total energy to three parameters.
However it is clear that calculating the total energy in the RPA+EXX scheme
removes the strong $U$-dependence observed in the PBE+U (and EXX) structural 
parameters, despite the implicit relation with $U$
through $\psi$ and $\e$.

\subsubsection{Relative stability of rutile and anatase phases}
\label{sec.TiO2_relstab}

An interesting property of TiO$_2$ is the competing stability
of the rutile and anatase polymorphs.
In nanostructured TiO$_2$ employed in photovoltaics,
anatase tends to be the dominant phase.\cite{Shklover1997} 
However the majority of experimental studies now agree
that in bulk crystalline TiO$_2$, rutile is more thermodynamically 
stable than anatase, with reported enthalpy differences ranging\footnote{We exclude a value
of 0.086 eV/f.u.\ listed in Ref.~\citenum{Ranade2002} due to its significant
(71\%) error bar.}
between 0.004 and 0.068 eV/formula unit (f.u.).\cite{Ranade2002}
Two recent experiments\cite{Ranade2002,Levchenko2006} found similar enthalpy differences of
0.027 and 0.017 eV/f.u.
The measurement of this quantity is a significant experimental challenge, requiring
careful control of impurity concentration and synthesis conditions.\cite{Ranade2002}

A number of theoretical works have calculated the relative total energies
of the anatase and rutile phases within DFT e.g.\ 
Refs.~\citenum{Dompablo2011,Curnan2015,Labat2007,Gerosa2015,
Muscat2002, Fahmi1993,Shirley2010, Conesa2010,Moellmann2012,Zhu2014}.
Approaches using LDA or GGA XC-functionals invariably determine
anatase to have a lower total energy than rutile.\cite{Labat2007,Muscat2002,Shirley2010}
Our own calculations using the PBE XC-functional and experimental geometries for the two
phases\cite{Burdett1987} reproduce this result, with an energy difference of
0.077~eV/f.u.; using optimized geometries slightly increases this value to 0.080~eV/f.u.
Inclusion of exact exchange through hybrid XC-functionals also predicts anatase
to have a lower energy,\cite{Curnan2015,Gerosa2015}
unless an unusually large amount ($>70$\%) of exact exchange is used.\cite{Curnan2015}

In common with most previous works, we note that our calculations
are missing the vibrational contribution
to the total energy; however the zero point contribution
was calculated to be only 0.01~eV/f.u.\ lower for rutile than anatase in 
Ref.~\citenum{Shirley2010}.\footnote{
Interestingly Ref.~\citenum{Moellmann2012} states that including the zero-point
motion should stabilize anatase, at variance with Ref.~\citenum{Shirley2010}}
However, it has been shown that rutile can be significantly stabilized with respect
to anatase within a DFT framework through two distinct routes, namely 
by adding either Hubbard $U$ terms to $H^0$ (GGA+$U$)\cite{Dompablo2011,Curnan2015}
or empirical corrections to account for dispersion interactions 
(DFT-D).\cite{Conesa2010,Moellmann2012,Zhu2014}
We note that even though both of these approaches can be used to obtain
the same qualitative result, they describe very different physics;
GGA+$U$ addresses strong, localized correlation, whilst DFT-D
attempts to capture relatively weak, long-range dispersion.
The advantage of our current RPA approach is that
it combines the Hubbard $U$ term with the
RPA description of long-range correlation.

\begin{figure}
\includegraphics{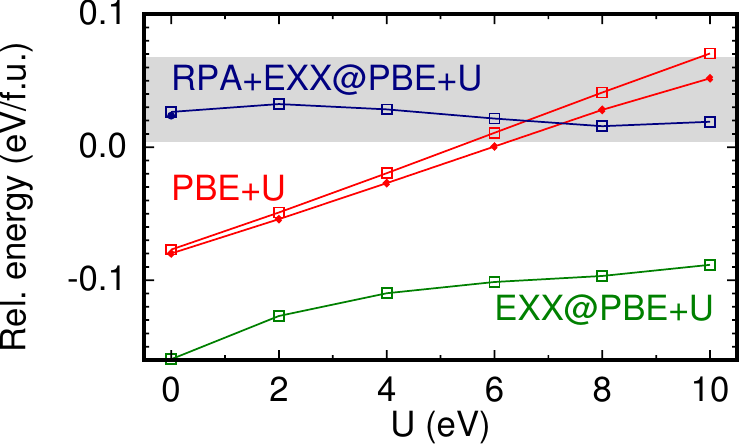}
\caption{(color online)
Total energy per formula unit 
of the anatase phase of TiO$_2$ given with respect to the rutile phase.
Open symbols denote calculations performed at the experimentally-measured
structures\cite{Burdett1987}, and filled symbols using the
structures optimized at the relevant level of theory.
The lines are guides to the eye.
We also illustrate the range of experimentally-measured enthalpy
differences between anatase and rutile (see text)\cite{Ranade2002}
as the gray shaded area.
\label{fig.TiO2_rel_stab}
}
\end{figure}
The red symbols in Fig.~\ref{fig.TiO2_rel_stab} show
the relative energies of anatase with respect to rutile 
within the PBE+$U$ approach.
The filled and empty symbols correspond to optimized
and experimental geometries, respectively.
In agreement with previous studies,\cite{Dompablo2011,Curnan2015}
we find that increasing the $U$ parameter stabilizes rutile,
with the two phases becoming energetically degenerate at $U\sim$5.5~eV.
The authors of Ref.~\citenum{Dompablo2011} further considered the columbite
phase of TiO$_2$, and noted that $U$ values in the range 5--8~eV
gave an energy ordering which matches the relative stability from
experiment.
Although these large $U$ values give band gaps close to experiment,\cite{Patrick2012}
they are somewhat larger than those calculated in Ref.~\citenum{Mattioli2008} 
or used e.g.\ in defect calculations.\cite{Dette2014}

Following the same approach as for the structural parameters,
we considered the difference between the anatase and rutile total energies 
calculated non-self-consistently including the EXX and RPA+EXX contributions.
The EXX calculations (green symbols in Fig.~\ref{fig.TiO2_rel_stab})
find anatase to have lower total energy than rutile regardless of the value
of $U$ used in the starting Hamiltonian.
This result is consistent with previous Hartree-Fock 
calculations.\cite{Labat2007,Muscat2002,Fahmi1993}
However the RPA+EXX calculations (blue symbols in Fig.~\ref{fig.TiO2_rel_stab}) 
show two interesting features:
First, even at $U=0$~eV, rutile has a lower energy than anatase, by
0.027~eV/f.u.
Second, increasing $U$ causes a non-monotonic variation in this difference 
only up to a maximum of 0.011~eV/f.u.
Thus regardless of the $U$ value used in the initial Hamiltonian, our calculated
non-self-consistent RPA total energy of rutile remains lower than that of anatase.

Since these RPA+EXX calculations were performed at the experimental
lattice parameters,\cite{Burdett1987} we checked the energy
difference obtained using RPA+EXX optimized structures\footnote{The experimental\cite{Burdett1987} 
(RPA+EXX optimized) lattice parameters for anatase TiO$_2$ at $U$=0~eV used were
$a$= 3.782~\AA \ (3.812~\AA), $c$= 9.502~\AA \ (9.567~\AA) and $u$ = 0.2083 (0.2083).}
for $U$=0~eV, and found a difference of only 0.003~eV/f.u. (filled blue symbol
in Fig.~\ref{fig.TiO2_rel_stab}).
This is the same difference observed between experimental and optimized
structures calculated within PBE+$U$ at $U$=0~eV.
The difference however is that the RPA optimized-structures depend less
strongly on $U$ than in PBE+$U$ (Fig.~\ref{fig.TiO2_structure}), so
we expect that using RPA+EXX optimized structures across the full $U$ range
to have an even smaller effect than that observed for the PBE+$U$ calculations.

Comparing our total energy calculations to the experimental enthalpy differences,
we find our calculations to lie within the experimental range (shaded area
of Fig.~\ref{fig.TiO2_rel_stab}).
We note that energy differences of $<$10~meV/f.u.\ lie 
at the limit of numerical accuracy currently achievable in our RPA calculations, 
and again emphasize that our calculations do not include vibrational contributions.
However by comparing the RPA+EXX and EXX total
energies in Fig.~\ref{fig.TiO2_rel_stab} it can be seen that the 
RPA correlation energy of rutile is more negative than that of anatase
by 0.186~eV/f.u.\ at $U$=0~eV, and by 0.108~eV at $U$=10~eV.
Therefore our calculations illustrate the key role played by non-local
correlation in understanding the phase stability of this material,\cite{Conesa2010}
and also demonstrate that the result is robust against
the choice of $U$ in $H^0(U)$.

\subsection{NiO}
\label{sec.NiO}
\subsubsection{Electronic structure}
The final material we consider is NiO, which in its 
paramagnetic state adopts a NaCl (F$m\overline{3}m$) structure.\cite{Cheetham1983}
Here we focus on the antiferromagnetic configuration formed
below the N\'eel temperature (523~K), where the spin direction
alternates between adjacent (111) Ni planes.
For simplicity we neglect the structural distortion
which accompanies this antiferromagnetic transition, since the deviation
from the cubic lattice is small ($<$0.1$^\circ$ angular variation
in lattice vectors).\cite{Cheetham1983}

\begin{figure}
\includegraphics{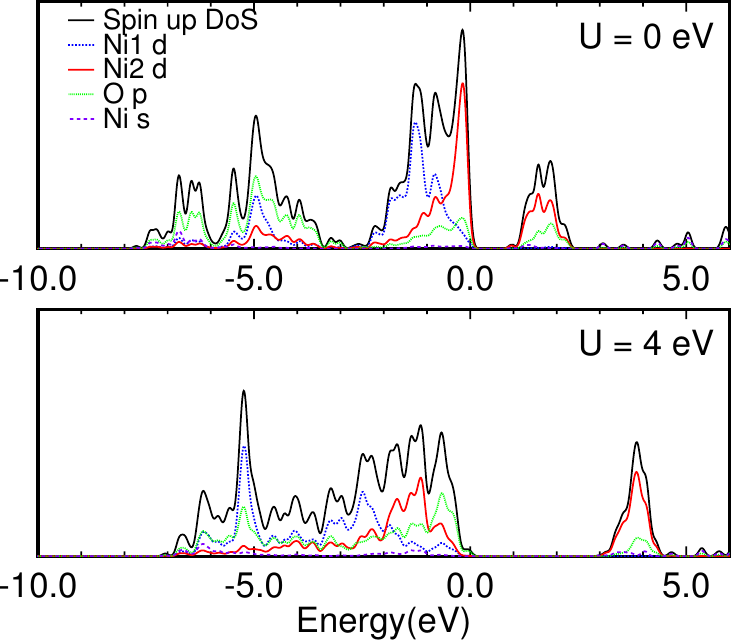}
\caption{(color online)
Projected-DoS calculated for NiO at the experimental lattice constant\cite{Cheetham1983}
using the PBE XC-functional with and without a $U$ correction of 4~eV.
We consider one spin direction, and use the labels Ni1 and Ni2 
to refer to the atoms with the majority of spins polarized
parallel and antiparallel to this direction, respectively.
\label{fig.NiO_pdos}
}
\end{figure}
In Fig.~\ref{fig.NiO_pdos} we show the NiO PDoS resolved for
one of the two spin components, calculated at the experimental
lattice constant\cite{Cheetham1983} (4.170~\AA) at the PBE+$U$ level for $U$=0 and 4~eV.
The PDoS demonstrates the complex character of the conduction
and valence bands, which both contain a substantial proportion of
Ni-3$d$ states.\cite{Cococcioni2005,Dudarev1998,Ansimov1991,Rohrbach2004}
The effect of the $U$ parameter is to open the gap between $d$ states,
which significantly increases the band gap from 1.0~eV at $U$=0~eV
to 3.0~eV at $U$=4~eV.
Furthermore the character of the band edges changes, such that the 
valence band edge is dominated by O-2$p$ states at $U$=4~eV (Fig.~\ref{fig.NiO_pdos}).
The ground-state spin-density is also strongly $U$-dependent, with the magnitude
of the local magnetic moment on the Ni atoms increasing from 1.4 to 1.8 Bohr
magnetons ($\mu_B$) over a $U$-range of 0--10~eV.\cite{Enkovaara2010}
It is also interesting to note that both the gap and local magnetic moment 
exhibit variation between the LDA (0.4~eV and 1.2$\mu_B$) and PBE (1.0~eV and 1.4$\mu_B$)
with $U$=0~eV.

\subsubsection{Atomistic structure}
\begin{figure}
\includegraphics{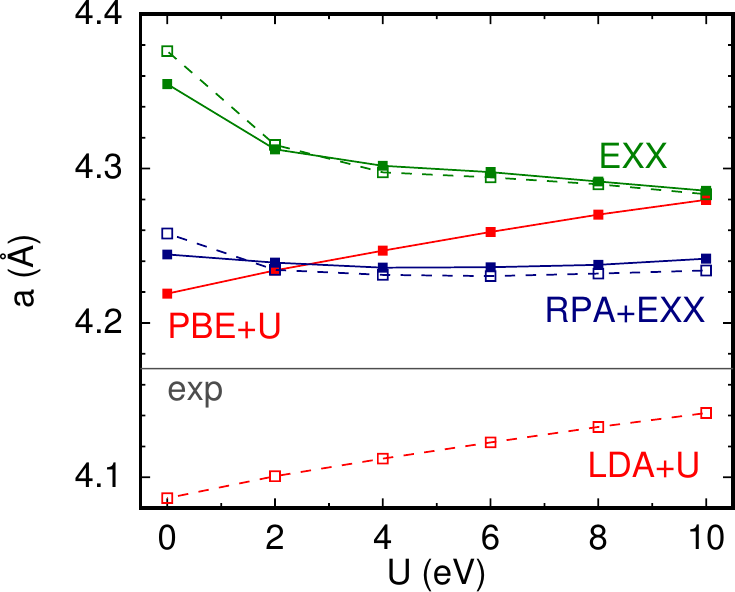}
\caption{(color online)
\label{fig.NiO_structure}
Lattice constant $a$ of NiO calculated under different
approximations (c.f.\ Fig.~\ref{fig.ZnS_structure} for labels).
Filled and empty symbols correspond to PBE+$U$ and LDA+$U$ calculations,
respectively.
The experimental lattice constant (gray horizontal line) 
was measured in Ref.~\citenum{Cheetham1983}.
}
\end{figure}
Given the strong $U$-dependence of the ground-state density, we
would expect the lattice constant of NiO also to be sensitive to $U$.
Figure~\ref{fig.NiO_structure} shows that this is indeed the case
when the total energy is obtained at the PBE+$U$ or LDA+$U$ level
(red symbols), with the lattice expanding for increased $U$.
Ref.~\citenum{Dudarev1998} noted that this expansion was accompanied
by a decrease in electronic charge in the interstitial regions, i.e.\
a reduction in covalent bonding.
Our calculated variation of LDA/PBE+$U$ lattice constants with $U$ in the range 
0--6~eV (0.04~\AA) is smaller than that reported in Ref.~\citenum{Dudarev1998}
(0.11~\AA) but larger than Ref.~\citenum{Rohrbach2004} ($<$0.01~\AA).
We attribute this difference to the frozen core approximation/core-valence
partitioning used in the PAW datasets.
The LDA and PBE calculations display the usual trend\cite{Patrick2015} of underestimating
and overestimating the experimental lattice constant\cite{Cheetham1983}
respectively, (-2.0\% and +1.1\% at $U$=0~eV).

The lattice constant calculated from $E^\mathrm{EXX}_\mathrm{Tot}$ 
with PBE wavefunctions overestimates the experimental 
value by 4.4\%.
This non-self-consistent value exhibits poorer agreement
with experiment than that obtained from Hartree-Fock 
calculations in Ref.~\citenum{Dudarev1998}, which overestimated the
experimental value by 2.1\%.
Initially, on including a $U$ correction of 2~eV  there is a relatively large 
decrease in lattice constant (0.04~\AA), but for higher $U$ values
the dependence is weaker ($<$0.03~\AA \ between $U$=2~and~10~eV).
Furthermore apart from a difference of 0.02~\AA \ at $U$=0~eV, using
LDA+$U$ wavefunctions to calculate $E^\mathrm{EXX}_\mathrm{Tot}$ 
yields very similar results to PBE+$U$ (green dashed lines in Fig.~\ref{fig.NiO_structure}).

The non-self-consistent RPA total energy calculations based on PBE+$U$
wavefunctions (blue solid line in Fig.~\ref{fig.NiO_structure}) 
overestimate the experimental lattice constant
by 1.6--1.7\% over the entire range of $U$ values.
The lattice constants obtained starting from LDA+$U$ (blue dashed lines) 
display the same trend as the EXX calculations, i.e.\ a larger
difference at $U$=0~eV compared to all other $U$ values.
In general the agreement with experiment is not as good as found
for the RPA calculations for TiO$_2$ and ZnS, and the PBE ($U$=0~eV)
lattice constant is closer to experiment.
Ref.~\citenum{Schimka2013} similarly found PBE to give a more accurate
lattice constant for elemental Ni than the RPA, with more recent work
attributing the difference to the quality of PAW datasets.\cite{Klimes20142}
However the most important feature of Fig.~\ref{fig.NiO_structure} is that,
like the other materials considered in this work, the non-self-consistent
RPA structural parameters are largely insensitive to the value of $U$
used in the initial Hamiltonian.
This perhaps is all the more remarkable for NiO, given the strong $U$-dependence
of the spin-density, band edge character and gap.

\subsection{$U$-dependence of total energy}
\label{sec.U_energies}
\begin{figure*}
\includegraphics{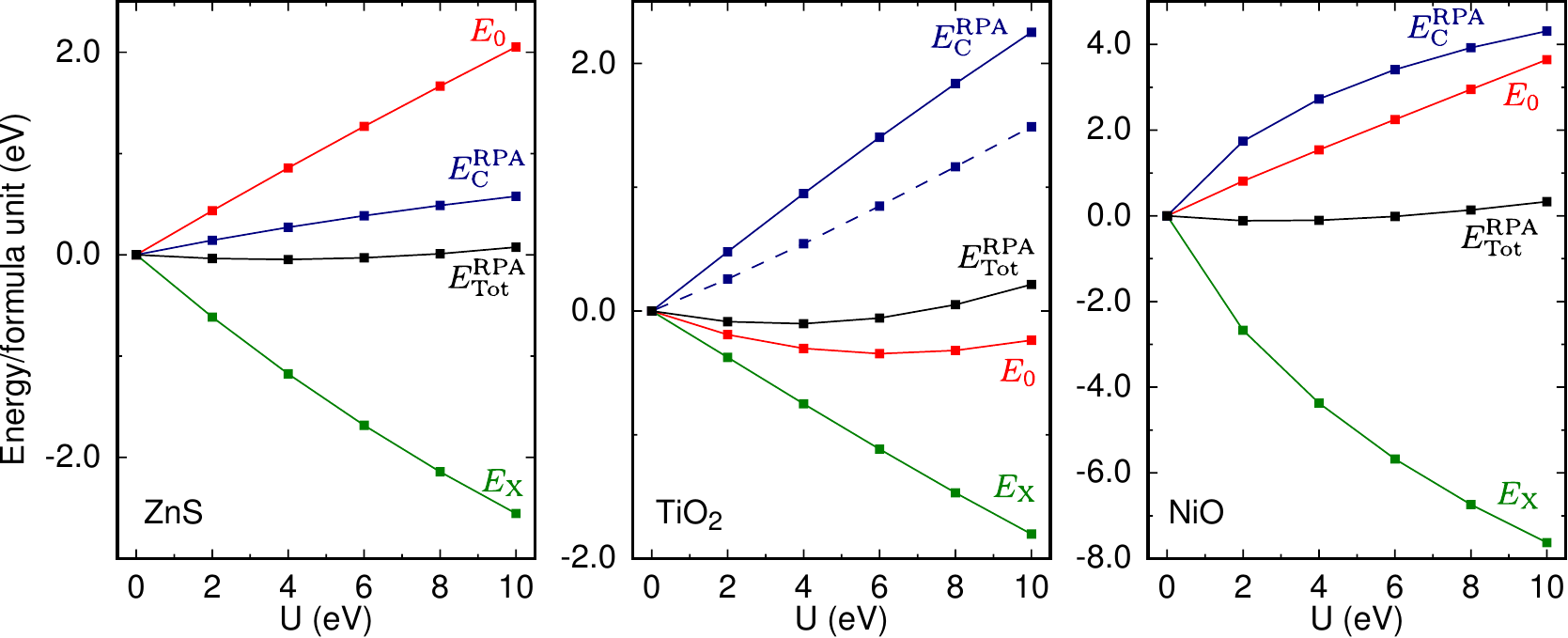}
\caption{(color online)
Decomposition of $E^\mathrm{RPA}_\mathrm{Tot}$ into its individual
contributions (equation~\ref{eq.etotRPA}) as a function of $U$ parameter
used in initial PBE+$U$ calculation.
Each quantity is given with respect to its calculated value at $U$=0~eV.
The dashed line shown for TiO$_2$ shows $E_\mathrm{C}^\mathrm{RPA}$
calculated with the effects of $U$ simulated with a scissor correction (see text).
\label{fig.energy_decomposition}
}
\end{figure*}
In order to further understand the effects of $H^0(U)$ on the calculated
value of $E^\mathrm{RPA}_\mathrm{Tot}$, in Fig.~\ref{fig.energy_decomposition}
we plot the individual contributions $E_0$, $E_\mathrm{X}$  and $E_\mathrm{C}^\mathrm{RPA}$
as a function of $U$ for each material at their experimental structures 
(for TiO$_2$ we show the results for the rutile phase).
The energies were calculated starting from PBE+$U$ wavefunctions, and the $U$=0~eV
value of each quantity has been used to define the energy zero.

The most notable aspect of Fig.~\ref{fig.energy_decomposition} is that
although $E_0$, $E_\mathrm{X}$  and $E_\mathrm{C}^\mathrm{RPA}$ are in general
strongly $U$-dependent (varying by several eV/f.u.\ over the considered $U$-range), 
the variation in their sum $E^\mathrm{RPA}_\mathrm{Tot}$ is an order of magnitude
smaller;
i.e.\ there is a strong cancellation between the $U$-dependent quantities.
In all cases, $E_\mathrm{X}$ becomes more negative with increasing $U$.
A simple explanation for this behavior is to 
note that the larger $U$ correction forces 
the electrons to occupy more atomic-like orbitals,
increasing the
self-interaction contribution to $E_\mathrm{X}$
(the $\nu_1=\nu_2$ term in equation~\ref{eq.Ex}).

The contributions which cancel $E_\mathrm{X}$ vary from material to material.
For ZnS, the $E_\mathrm{X}$ contribution is mainly balanced by $E_0$,
whilst for TiO$_2$ it is $E_\mathrm{C}^\mathrm{RPA}$.
In NiO both $E_0$ and $E_\mathrm{C}^\mathrm{RPA}$ contribute.
The behavior of $E_0$ with $U$ depends on whether the $3d$ states are occupied
(ZnS, NiO) or mainly unoccupied (TiO$_2$).
In the former case, the $U$ term causes the $3d$ states to become more 
localized, which carries a kinetic energy penalty and thus increases $E_0$.
By contrast for TiO$_2$, the $U$ correction depopulates the $3d$ states and pushes
these electrons into the less-localized $2p$ orbitals, reducing
the kinetic contribution.

The RPA correlation energy $E_\mathrm{C}^\mathrm{RPA}$ becomes more
positive (i.e.\ decreases in magnitude) with increasing $U$.
The principal cause of this behavior is the increase in band gap,
which reduces the screening through the energy denominators
in $\chi_\mathrm{KS}$ (equation~\ref{eq.chiks}).
The increased variation of $E_\mathrm{C}^\mathrm{RPA}$ across
ZnS$\rightarrow$TiO$_2$$\rightarrow$NiO reflects the sensitivity
of the material's band gap to $U$.
However, the observed behavior of $E_\mathrm{C}^\mathrm{RPA}$  
cannot be viewed entirely in terms of the band gap.
To illustrate this point, in Fig.~\ref{fig.energy_decomposition} for
TiO$_2$ we show the correlation energy calculated using the PBE ($U$=0~eV)
wavefunctions, where the effect of $U$ on the band gap was mimicked by 
applying a scissor correction to the unoccupied states used to construct
$\chi_\mathrm{KS}$.
Specifically, the size of the scissor correction was related to $U$
through Fig.~\ref{fig.TiO2_bs_gap}(c) to reproduce the $\Gamma$-$\Gamma$ 
gap.
As shown by the dashed line in Fig.~\ref{fig.energy_decomposition},
the scissor-correction accounts for $\sim$65\% of the variation
in $E_\mathrm{C}^\mathrm{RPA}$.
In order to account for the remaining 35\% it is therefore necessary 
to also consider the $U$-dependent variations 
of the bandstructure (e.g.\ the position of the $e_g$ and $t_{2g}$ subbands) 
and the shapes of the wavefunctions.

\begin{figure*}
\includegraphics{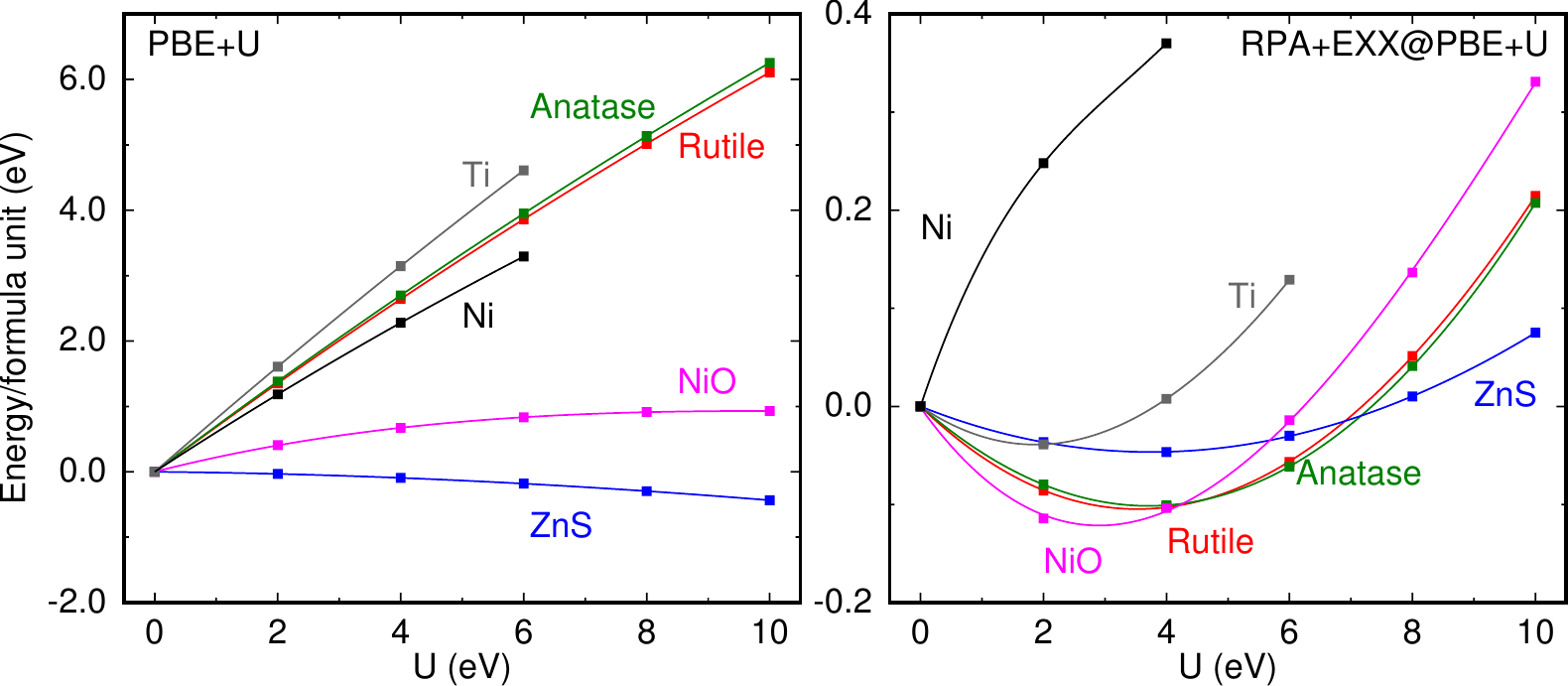}
\caption{(color online)
\label{fig.energy_comparison}
Total energies per formula unit as a function of $U$, calculated 
(left) self-consistently from the PBE+$U$ XC-functional, and 
(right) non-self-consistently from equation~\ref{eq.etotRPA}.
Both the rutile and anatase TiO$_2$ polymorphs are considered.
Each quantity is given with respect to its calculated value at $U$=0~eV.
The lines are polynomial fits to the calculated data points (squares).
}
\end{figure*}
In Fig.~\ref{fig.energy_comparison} we compare the magnitude of
variation of $E^\mathrm{RPA}_\mathrm{Tot}$ with the self-consistent
total energy obtained with the PBE+$U$ XC-functional.
The metals Ti and Ni are included in this analysis; these calculations 
are discussed in more detail in Section~\ref{sec.heat_form} below.
Comparison of the scales on the $y$-axis emphasizes how the self-consistent PBE+$U$
energy is much more sensitive to $U$ than $E^\mathrm{RPA}_\mathrm{Tot}$.
In the case of TiO$_2$, this difference is a factor of 30.
A further interesting point regarding TiO$_2$ is the energy difference
between the anatase and rutile polymorphs shown in Fig.~\ref{fig.TiO2_rel_stab}.
Here we see that the variation in energy difference between the two polymorphs,
going from -0.08~to~0.07~eV/f.u.\ over the $U$ range of 0--10~eV,
is 40 times smaller than the variation in the self-consistent PBE+$U$ 
energy of each phase.
By contrast the variation in $E^\mathrm{RPA}_\mathrm{Tot}$
is the same order of magnitude as the energy difference.

\subsection{Minimization of $E^\mathrm{RPA}_\mathrm{Tot}$ with $H^0(U)$}
\label{sec.U_min}
\begin{table}
\caption{
\label{tab.U_vals} 
Values of $U_\mathrm{min}$ obtained for the materials considered
in this work.
$U_\mathrm{min}$ is determined from Fig.~\ref{fig.energy_comparison}
as the $U$ value at which each curve is at a minimum.
We compare our results to $U$ values reported from previous calculations.
}
\begin{tabular}{lcc}
\hline
\hline
                 &  $U_\mathrm{min}$ (eV)& Previously calculated $U$ (eV)\\
\hline
ZnS              &   3.7                 & 6.0\footnotemark[1], 7.0\footnotemark[2]\\
TiO$_2$ (rutile) &   3.5                 & 3.4\footnotemark[3], 6.0\footnotemark[4] \\
TiO$_2$ (anatase)&   3.7                 & 3.3\footnotemark[3], 5.3\footnotemark[4], 7.5\footnotemark[5] \\
NiO              &   2.9                 & 7.1,\footnotemark[6] 4.6,\footnotemark[7], 6.2,\footnotemark[8] 6.4\footnotemark[9]\\
Ti               &   1.8                 & ---\\
Ni               &   ---                 & --- \\
\hline
\hline
\footnotetext[1]{$U_H -J $, constrained DFT, Ref.~\citenum{Jiang2010}}
\footnotetext[2]{Matrix elements of screened Coulomb interaction, Ref.~\citenum{Miyake2006}}
\footnotetext[3]{Linear response formalism, Ref.~\citenum{Mattioli2008}}
\footnotetext[4]{Linear response formalism, Ref.~\citenum{Curnan2015}}
\footnotetext[5]{Matching of $G_0W_0$ and PBE+$U$ band gap, Ref.~\citenum{Patrick2012}}
\footnotetext[6]{Constrained DFT, Ref.~\citenum{Ansimov1991}}
\footnotetext[7]{Linear response formalism, Ref.~\citenum{Cococcioni2005}}
\footnotetext[8]{Fit to experimental electron energy loss spectrum, Ref.~\citenum{Dudarev1998}}
\footnotetext[9]{Fit to experimental heat of formation, Ref.~\citenum{Wang2006}}
\end{tabular}
\end{table}
Interestingly, Fig.~\ref{fig.energy_comparison} also demonstrates
that it is possible to minimize $E^\mathrm{RPA}_\mathrm{Tot}$ 
with respect to the continuum of single-particle Hamiltonians $H^0(U)$
defined by $U$,
and thus introduce a material-dependent quantity $U_\mathrm{min}$
at which $E^\mathrm{RPA}_\mathrm{Tot}$ is a minimum.
It is shown in Ref.~\citenum{IsmailBeigi2010} that a blind optimization
of $E^\mathrm{RPA}_\mathrm{Tot}$ with respect to all possible $H^0$ (where
$H^0$ contains a nonlocal potential) will push all eigenvalues to the
Fermi level and thus cause $E^\mathrm{RPA}_\mathrm{Tot} \rightarrow -\infty$.
In the same work it is suggested that a sensible method of proceeding
is to somehow constrain $H^0$ so as to avoid this unphysical behavior.
The current work can be seen as an implementation of this idea, where
specifically we have restricted our search to Hamiltonians of the form $H^0(U)$
(equation~\ref{eq.spH}).

In Table~\ref{tab.U_vals} we compare our obtained $U_\mathrm{min}$
to other values of $U$ used in previous works.
Although it is less common to apply $U$ corrections to metals,\cite{Ansimov19912}
standard PBE+$U$ calculations of 
heats of formation find it necessary to apply the Hubbard $U$ also
to the metallic system, with reasonable results.\cite{Yan2013}
Our numbers are generally smaller than those used in other works; of course
given our unique criterion of determining $U_\mathrm{min}$, there is
no reason why they should agree.
Indeed the value of $U$ depends on the choice made for the projector functions,\cite{Pickett1998}
and can vary on the scale of electronvolts depending on the 
treatment of the core-valence interaction.\cite{Curnan2015}

We note that the values of $U_\mathrm{min}$ obtained here give
reasonable physical properties, such as a local magnetic moment
of 1.6~$\mu_B$ for NiO (experimental values range from 
1.6--1.9~$\mu_B$).\cite{Cococcioni2005}
However it is also true that the computational cost of obtaining
$U_\mathrm{min}$ does not make the above scheme an
attractive method of selecting $U$ compared to other methods.\cite{Ansimov1991,Cococcioni2005}
Indeed the quite weak sensitivity of the energy to 
the value of $U$ combined with the numerical uncertainty inherent 
in such calculations means that we must attach caution to the
values listed in Table~\ref{tab.U_vals}.
Nonetheless it would be interesting to explore
the minimization of $E^\mathrm{RPA}_\mathrm{Tot}$ with
respect to $H^0(U)$ for an extended range of TMCs.

\subsection{Heats of formation of TiO$_2$ and NiO}
\label{sec.heat_form}

A recent work\cite{Yan2013} presented calculations of the heats of formation
for a range of oxides, allowing comparison of the performance of different
total energy methods, including the non-self-consistent RPA.
TiO$_2$ (rutile) and NiO were among the materials considered in Ref.~\citenum{Yan2013}, and
are notable because of the very good (TiO$_2$) and very poor (NiO) agreement found
between their calculated heats of formation and experiments.
To make contact with that work, we also calculated the heats
of formation, obtained per oxygen atom as
\begin{equation}
\Delta E_\mathrm{O} = \frac{1}{y} E(\mathrm{A}_x\mathrm{O}_y) - \frac{x}{y}E(\mathrm{A}) - \frac{1}{2}E(\mathrm{O}_2)
\label{eq.hof}
\end{equation}
where $E(\mathrm{A}_x\mathrm{O}_y)$, $E(\mathrm{A})$
and $E(\mathrm{O}_2)$ are the energies per formula unit of the oxide, metal and oxygen molecule respectively.
We used experimental lattice parameters throughout, with Ti in a hcp
structure (P$6_3/mmc$), $a$=2.957~\AA \ and $c/a$=1.585,\cite{Vohra2001}
and (ferromagnetic) Ni in a fcc structure (F$m\overline{3}m$) with $a$=3.516~\AA.\cite{Haas2009}

\begin{figure}
\includegraphics{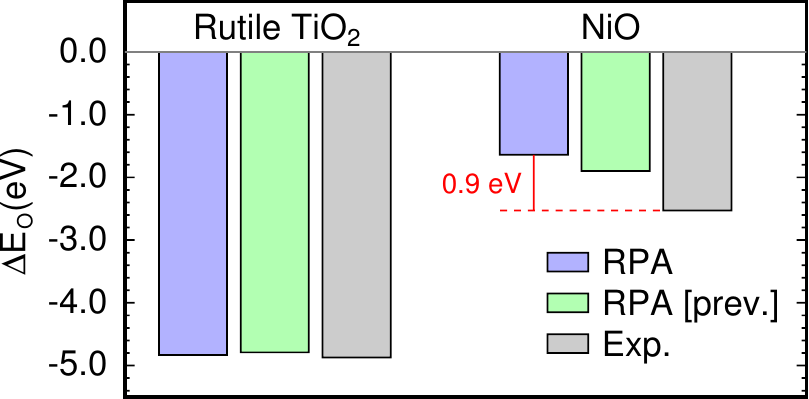}
\caption{(color online)
Heat of formation per oxygen atom $\Delta E_\mathrm{O}$ calculated from
equation~\ref{eq.hof}.
Blue and green bars represent non-self-consistent RPA total energy
calculations performed using PBE wavefunctions in the current
and previous (prev.) work.\cite{Yan2013}
Gray bars represent experimental values.\cite{Yan2013}
\label{fig.heat_formation}
}
\end{figure}
The values of $\Delta E_\mathrm{O}$ for TiO$_2$ and NiO calculated from 
$E^\mathrm{RPA}_\mathrm{Tot}$ (PBE wavefunctions, $U$=0~eV)
are presented in Fig.~\ref{fig.heat_formation}.
We compare our results to the calculations 
and room temperature experimental values reported in Ref.~\citenum{Yan2013}.
Focusing first on the calculations,
we find good agreement (0.04~eV) between our $\Delta E_\mathrm{O}$
and that of Ref.~\citenum{Yan2013} for TiO$_2$ .
However there exists a difference of 0.3~eV in $\Delta E_\mathrm{O}$ for NiO,
which we assign to our explicit treatment of the Ni 3$p$ states.
If instead these states as frozen in the Ni core we obtain a value of 
$\Delta E_\mathrm{O}$ of $-1.84$~eV, much closer to the $-1.90$~eV
reported in Ref.~\citenum{Yan2013}.
We also note that Ref.~\citenum{Yan2013} used PBE structural parameters, whilst
here we use experimental values; this aspect also explains the difference
in $\Delta E_\mathrm{O}$ for rutile TiO$_2$ calculated here and in
Ref.~\citenum{Jauho2015}.

Now considering experiment,
for rutile TiO$_2$ there is close agreement with the non-self-consistent
RPA with a difference in $\Delta E_\mathrm{O}$ of 0.04~eV.
However as emphasized in Fig.~\ref{fig.heat_formation}, for NiO there is a significant 
discrepancy (0.9~eV), with the non-self-consistent RPA apparently underestimating
the stability of NiO compared to Ni.
Ref.~\citenum{Yan2013} found similarly poor performance for the monoxides VO and CoO,
and Cr$_2$O$_3$.

In the context of the current work it is natural to ask whether one
can obtain RPA values of $\Delta E_\mathrm{O}$ closer to experiment
by including a $U$ correction in the initial Hamiltonian.
This approach can be tested immediately from the data shown in Fig.~\ref{fig.energy_comparison}.
Choosing the $U$ value as $U_\mathrm{min}$ would shift $\Delta E_\mathrm{O}$ to more
negative values for both TiO$_2$ and NiO.
For TiO$_2$ the new $\Delta E_\mathrm{O}$ is 0.03~eV lower in energy, essentially
reproducing the experimental value (although no vibrational effects were taken
into account in the calculations).
For NiO, the correction is -0.12~eV which, although slightly reducing the discrepancy
with experiment, does not account for the 0.9~eV difference.

NiO has long been recognized as a system representing a major challenge
to density-functional based methods,\cite{Ansimov1991} and we also note that metallic Ni cannot
be considered straightforward either.\cite{Starrost2001}
One option is to go beyond the RPA in the calculation of the correlation
energy, for instance through the introduction of a time-dependent DFT
kernel in the integral equation for $\chi^\lambda(\omega)$.\cite{Olsen2013}
Recently it was found that such an approach employing a static kernel
based on the homogeneous electron gas reduced the absolute error in $\Delta E_\mathrm{O}$
by 0.2~eV for a range of metal oxides, compared to the RPA.\cite{Jauho2015}
Further exploration of kernels which have a frequency dependence or display a small-wavevector
divergence\cite{Patrick2015} would be an interesting direction for future study.

\section{Conclusions}
\label{sec.Conclusions}

We have presented a study into the effects 
of including a Hubbard $U$ correction in the calculation of the single-particle
wavefunctions used to construct the non-self-consistent exact exchange and
RPA correlation energy.
We have explored materials where the 3$d$ band is fully occupied (ZnS), almost
empty (TiO$_2$) and partly occupied (NiO), and determined the $U$-dependence
of their lattice constants.
We have further addressed the question of the relative stability of the TiO$_2$
polymorphs anatase and rutile, and the heats of formation of the oxides TiO$_2$ and NiO.

The principal conclusion of this work is that the lattice constants derived 
from the non-self-consistent
RPA total energy $E^\mathrm{RPA}_\mathrm{Tot}$
are remarkably robust against changes to the value of $U$
in the starting Hamiltonian.
NiO is a good example: Including a $U$ correction opens the band
gap, redistributes the spin density and changes the character of the
band edges, yet the non-self-consistent RPA lattice constant changes
by less than 0.01~\AA \ over $U$ values ranging from 0--10~eV.

We have further shown that $E^\mathrm{RPA}_\mathrm{Tot}$ itself
is far less sensitive to $U$ than the self-consistent PBE+$U$
total energy.
This insensitivity originates from competing $U$-dependences
of the non-interacting ($E_0$), exchange ($E_\mathrm{EXX}$) 
and correlation  ($E_\mathrm{C}^\mathrm{RPA}$) energies.
For the materials considered here we have shown it is possible
to minimize $E^\mathrm{RPA}_\mathrm{Tot}$ with respect
to the $U$ value by choosing the single-particle Hamiltonian
$H^0(U=U_\mathrm{min})$.

For the specific case of TiO$_2$, we have found the
difference in $E^\mathrm{RPA}_\mathrm{Tot}$ between 
rutile and anatase polymorphs to vary
by less than 0.01~eV per formula unit over the entire $U$ range.
This variation is an order of magnitude smaller than that
calculated self-consistently at the PBE+$U$ level.
Furthermore, the non-self-consistent RPA energy ordering
reflects the ordering of experimental enthalpies.

The observed insensitivity of $E^\mathrm{RPA}_\mathrm{Tot}$ to $H^0$
should be considered a positive attribute of non-self-consistent
RPA total energy calculations of the structural properties
of solids,  and distinguishes the method from
$G_0W_0$ calculations of quasiparticle energies which display
a stronger starting point dependence.
By the same token however, situations which are
problematic for the RPA based on GGA or LDA Hamiltonians
are unlikely to be improved by attaching a $U$ correction to $H^0$.
We have demonstrated this explicitly in the case of the heat of formation 
of NiO, where the inclusion of $U$ corrections can only reduce
the discrepancy with experiment by a small amount.
Such cases must therefore remain a challenge for beyond-RPA methods.

\begin{acknowledgments}
We acknowledge support from the Danish Council for 
Independent Research's Sapere Aude Program, Grant No. 11-1051390.
\end{acknowledgments}
%
\end{document}